\documentclass[aps,prl,twocolumn,superscriptaddress]{revtex4}

\usepackage{mathrsfs,amssymb,graphics,subfigure,threeparttable}
\usepackage{grap hicx}
\usepackage{amssymb}
\usepackage{enumerate}
\usepackage{amsmath}
\usepackage{fancyhdr}
\usepackage{bm}
\usepackage[squaren]{SIunits}

\begin{document}

%\preprint{}

\title{Nano-scale electron bunching in laser-triggered ionization injection in plasma accelerators}


\author{X. L. Xu}
\affiliation{Department of Engineering Physics, Tsinghua University, Beijing 100084, China}
\affiliation{University of California, Los Angeles, California 90095, USA}
\author{C. J. Zhang}
\affiliation{Department of Engineering Physics, Tsinghua University, Beijing 100084, China}
\author{F. Li}
\affiliation{Department of Engineering Physics, Tsinghua University, Beijing 100084, China}
\author{Y. Wan}
\affiliation{Department of Engineering Physics, Tsinghua University, Beijing 100084, China}
\author{Y. P.  Wu}
\affiliation{Department of Engineering Physics, Tsinghua University, Beijing 100084, China}
\author{J. F.  Hua}
\affiliation{Department of Engineering Physics, Tsinghua University, Beijing 100084, China}
\author{C.-H.  Pai}
\affiliation{Department of Engineering Physics, Tsinghua University, Beijing 100084, China}
\author{W. Lu}
\email[]{weilu@tsinghua.edu.cn}
\affiliation{Department of Engineering Physics, Tsinghua University, Beijing 100084, China}
\author{W. An}
\affiliation{University of California, Los Angeles, California 90095, USA}
\author{P. Yu}
\affiliation{University of California, Los Angeles, California 90095, USA}
\author{W. B. Mori}
\affiliation{University of California, Los Angeles, California 90095, USA}
\author{C. Joshi}
\affiliation{University of California, Los Angeles, California 90095, USA}

\date{\today}

\begin{abstract}
Ionization injection is attractive as a controllable injection scheme for generating high quality electron beams using plasma-based wakefield acceleration. Due to the phase dependent tunneling ionization rate and the trapping dynamics within a nonlinear wake, the discrete injection of electrons within the wake is nonlinearly mapped to discrete final phase space structure of the beam at the location where the electrons are trapped. This phenomenon is theoretically analyzed and examined by three-dimensional particle-in-cell simulations which show that three dimensional effects limit the wave number of the modulation to between $> 2k_0$ and about $5k_0$, where $k_0$ is the wavenumber of the injection laser. Such a nano-scale bunched beam can be diagnosed through coherent transition radiation upon its exit from the plasma and may find use in generating high-power ultraviolet radiation upon passage through a resonant undulator.
\end{abstract}

\pacs{}
%\keywords{}

\maketitle

%\section{\label{sec: Introduction}Introduction}

\begin{figure}[bp]
\includegraphics[width=0.5\textwidth]{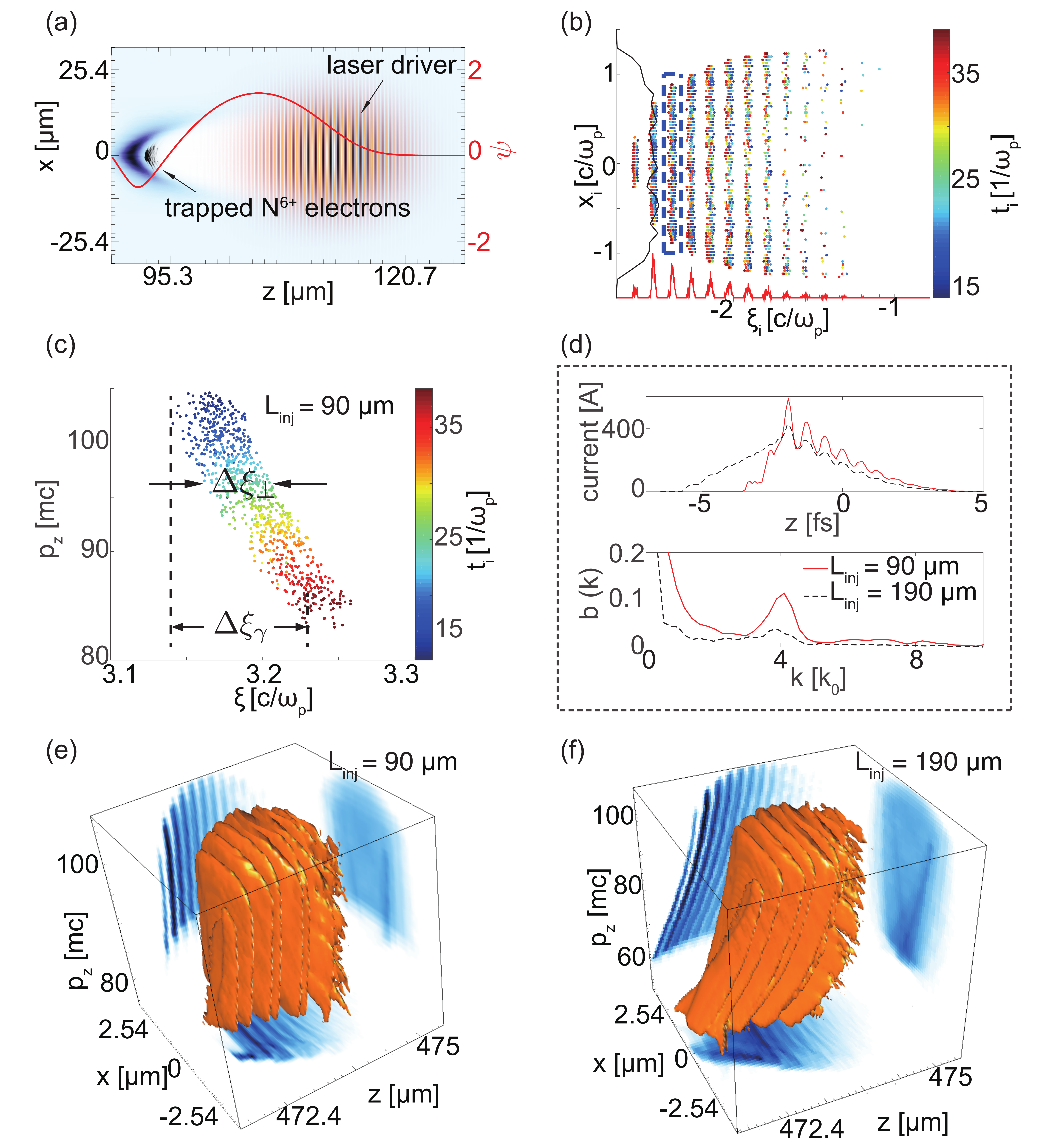}
\caption{\label{fig: } The nano-scale bunching of injected charge in a PWFA.  The laser is focused at $z=0~\milli\meter$ plane. (a) Snapshot of the charge density distribution of the background electrons, the K-shell electrons of nitrogen, and the laser electric field in $x$ direction. The red line is the on-axis ($x=y=0$) pseudo potential $\psi$. (b) The distribution of the ionization injected electrons in $(\xi_i, x_i)$ space. The color represents the time when the electrons are ionized. The red peaks are integrated injected charge in $x_i$ whereas the black is the integrated injected charge for all $\xi_i$. (c) The $(\xi, p_z)$ space at $z=0.5~\milli\meter$ for an initial slice indicated by the dashed box in (b). (d) The current profile and the bunching factor at $z= 0.5~\milli\meter$. (e) and (f) show the $(z, x, p_z)$ phase space at $z= 0.5~\milli\meter$. For $L_{inj}=90~\micro\meter$ case, the $\mathrm{N}^{5+}$ ions are distributed from $z=0.038~\milli\meter$ to $z=0.128~\milli\meter$ and $n_{\mathrm{N}^{5+}}=10^{-3}n_p$; for $L_{inj}=190~\micro\meter$ case, the $\mathrm{N}^{5+}$ ions are distributed from $z=0.038~\milli\meter$ to $z=~0.228\milli\meter$ and $n_{\mathrm{N}^{5+}}=5\times10^{-4}n_p$. }
\end{figure}

Due to the ability to sustain ultra-high acceleration gradients ($\giga\volt\per\centi\meter$), the field of plasma-based wakefield acceleration has attracted much attention in the past two decades \cite{joshi2003plasma}. Recently, ionization injection has been proposed and demonstrated \cite{chen2006electron, PhysRevLett.98.084801, PhysRevLett.104.025003, PhysRevLett.104.025004, PhysRevLett.105.105003, PhysRevLett.107.035001, PhysRevLett.107.045001, PhysRevLett.112.025001} as a viable injection scheme and investigated for generating high brightness ($\sim 10^{19} \ampere/\rad^{2}/\meter^{2}$), stable, and tunable electron beams \cite{PhysRevLett.108.035001, PhysRevLett.111.015003, PhysRevLett.111.155004, PhysRevLett.111.245003, PhysRevLett.112.125001, PhysRevSTAB.17.061301}. The basic idea is that the trapping threshold of an electron is reduced when it is born inside the wakefield near the maximum of wake's potential compared to an electron from a pre-ionized plasma. Such high brightness beams are needed for future free-electron-laser and collider applications. 

The key to generating a high brightness beam is to limit the volume within the wake where injection of electrons occurs \cite{PhysRevLett.112.035003}. In the case where injection is from field ionization due to a laser pulse, the ionization volume is limited by choosing the intensity of the injection pulse(s) close the ionization threshold of bound electrons. Therefore, electrons are mostly born near the peaks and the troughs of the oscillating laser electric field. The phase-dependent ionization leads to an intrinsic initial phase space discretization at twice the optical frequency, which is known to produce third harmonic generation in tunnel ionized plasma \cite{PhysRevLett.68.321, PhysRevA.46.1091}. 
 
We show in this Letter using theory and fully three-dimensional (3D) particle-in-cell (PIC) simulations, that when ionization occurs on either side of the peak of the wake potential the electron bunch can be strongly modulated in space on the nano-meter scale when it becomes trapped. In the 1D limit the spacing of the modulations can be made arbitrarily small. However, we show that three-dimensional effects limit the discretization pattern to less than one-fifth the laser wavelength. The concept is robust and has the potential to provide lower overall energy spread, lower emittances, shorter modulation wavelengths, and more nano bunches than another recently proposed scheme \cite {PhysRevLett.114.084801}. Such an ultra-short and micro-bunched electron beam can be diagnosed via the coherent transition radiation upon exiting the plasma \cite{leemans2003observation} and may be used to produce high power coherent EUV radiation in a short resonant undulator. 

To illustrate the concept, we first consider ionization injection using a single laser pulse as shown in Fig. 1(a). An 800 $\nano\meter$ laser pulse polarized in the ${x}$ direction with normalized vector potential $a_0=2, w_0=14~\micro\meter$ and a pulse length (fwhm of energy) of 26 fs, propagates into a mixture of pre-ionized plasma and $\mathrm{N}^{5+}$ ions. The pre-ionized electrons form a nonlinear wake. As has been observed previously \cite{PhysRevLett.104.025003}, the K-shell electrons of nitrogen with high IPs are released during the rising edge of the wake potential [Fig. 1(a)], then slip to the back of the wake where some of these electrons are trapped \cite{PhysRevLett.104.025003}. This process is examined using the 3D PIC code OSIRIS \cite{fonseca2002high} using a moving window \cite{PhysRevLett.72.490}. We define the $z$ axis to be the laser propagating direction. The code uses the Ammosov-Delone-Krainov (ADK) tunneling ionization model \cite{ADKionizationmodel1986}. The IPs of the sixth and seventh nitrogen electrons are $I_p \approx 552.1~\electronvolt, 667.0~\electronvolt$ respectively, and the Keldysh parameter in this simulation is $\gamma_K= \sqrt{I_p/2U_p} \approx 0.023, 0.021 \ll 1$; therefore the ADK model should be valid. The simulation window has a dimension of $63.5 \times 63.5 \times 38.1~\micro\meter$ with $500\times 500\times 1500$ cells in the $x, y$ and $z$ directions, respectively. This corresponds to cell sizes of $k_0^{-1}$ in the $x$ and $y$ directions and 0.2 $k_0^{-1}$ in the $z$ direction, where $k_0$ is the wavenumber of the laser pulse.  

The $(\xi_i, x_i)$ space distribution of the trapped electrons when they are ionized is shown in Fig. 1(b), where $\xi \equiv v_\phi t-z$ is the relative longitudinal position and $v_\phi$ is the phase velocity of the wake. Due to the laser phase-dependent ionization probability, the initial electron distribution has a strong modulation at $2k_{0}$. After being released, the electrons slip to the back of the wake and are accelerated by the longitudinal electric field in the wake. Under the quasi-static approximation, $\gamma - (v_\phi/c) p_z - \psi =\mathrm{Const}$ \cite{mora1997kinetic}, where $p_z$ is normalized to $mc$, $\psi \equiv (e/mc^2) [\phi - (v_\phi /c) A_z]$ is the pseudo potential, $\psi$ in the fully blown-out wake can be expressed as $\psi \approx [r_b^2(\xi) - r^2] /4$ \cite{PhysRevLett.96.165002}\cite{lu2006nonlinearPoP}. Here $r_b (\xi)$ is the normalized radius of the ion channel that has a spherical shape for a sufficiently large maximum blowout radius $r_m$ given by $r_b^2(\xi) = r_m^2 - \xi^2 $ \cite{PhysRevLett.96.165002}\cite{lu2006nonlinearPoP}. Note that all parameters with units of length are normalized to the background plasma skin depth. Using the constant of motion given above, the relative longitudinal position of the injected electron can be expressed as
\begin{align}
\xi \approx \sqrt{4+\xi_i^2+r_i^2-r^2 - 4\left[\gamma - (v_\phi/c) p_z\right]}
\end{align}
 The electron conducts betatron oscillations in $x$ and $y$ with a decreasing amplitude under the focusing and acceleration fields \cite{PhysRevLett.112.035003}\cite{PhysRevLett.88.135004}. An initial isolated slice in $(\xi_i, x_i, y_i)$ will be mapped to an isolated slice in $(\xi, x, y, p_z)$ space. If $r_i \ll 1$ and the transverse momentum (the vector potential of the laser at the time of ionization) are small, and the electrons are relativistic (i.e., $\gamma - (v_\phi/c)p_z \ll 1$), the position $\xi$ is mainly determined by the initial $\xi_i$ as $\xi \approx \sqrt{4+\xi_i^2}$, which means the initial modulation in $\xi_i$ can be nonlinearly mapped to $\xi$. This means that an initial slice (electrons with the same $\xi_i$) will be mapped to the same final slice (same $\xi$).

Any spread in $r_i$, $r$ and $\gamma-(v_\phi/c)p_z$ will broaden the $\xi$ distribution for an initial slice. This can be seen in the simulation results shown in Fig. 1. In Fig. 1(c) we show the $(\xi, p_z)$ space at $z=0.5~\milli\meter$ for an initial slice [indicated by the dashed box in Fig. 1(b)]. One can see that the spread of $\xi$ due to the spread of transverse motion is $\Delta\xi_\perp \approx 0.05$. For this example where a laser driver with moderate $a_0$ is used, we find that, $1-v_\phi/c \approx \omega_p^2/(2\omega_0^2)$ \cite{mori1997IEEE}. Therefore,  the term $\gamma-(v_\phi /c) p_z\approx \gamma(1-v_\phi/c)\approx \omega_p^2/(2\omega_0^2) \gamma$ contributes differently for electrons with different energy leading to a spread in  $\xi$ for electrons with the same $\xi_i$.  Specifically, electrons ionized earlier (at different $z_i$) but at the same $\xi_i$ can have higher energy and smaller $\xi$. In Fig. 1(c) the difference in $\xi$ due to the energy difference is seen to be $\Delta\xi_\gamma \approx 0.1 $. This spread depends on the spread in $z_i$ which can be controlled by limiting the duration (distance) of ionization, $L_{inj}$. In the simulations we increased  $L_{inj}$ from $90\micro\meter \approx 24 c/\omega_p$ to $190\micro\meter \approx 50 c/\omega_p$, by varying the region where $\mathrm{N}^{5+}$ existed. In Figs. 1(d)-(f) it can be seen that the difference of $\xi$ due to this spread in energy is increased to $\Delta\xi_\gamma \approx 0.2$. The current profile and the bunching factor (defined as $b(k) = \left| \int dz g(z) \mathrm{exp}(i k z) \right|$, where $g(z)$ is the normalized distribution of the trapped electrons) are shown in Fig. 1(d). The modulation in the current profile is peaked at $k\approx 4k_0$, and the modulation and the bunching factor are reduced when $L_{inj}$ is increased from 90 $\micro\meter$ to 190 $\micro\meter$ due to the larger $\Delta\xi_\gamma$. The discretized phase space structure can be seen clearly in $(z,x,p_z)$ phase space at $z=0.5\milli\meter$ as shown in Fig. 1(e) and (f), however, for the larger $L_{inj}$, the slices are slanted in $(p_z, z)$ space indicating that within a narrow energy slice of the beam the bunching factor can still be large. 

By using two pulses to separate the wake formation and the electron injection, the initial and final phase space of the trapped electrons can be better controlled \cite{PhysRevLett.108.035001, PhysRevLett.111.015003, PhysRevLett.111.245003, PhysRevLett.112.125001, PhysRevLett.112.035003, PhysRevSTAB.17.061301}. Throughout the rest of this Letter, we consider the driver pulse to be a relativistic electron bunch and the injection pulse to be a co-propagating low intensity laser pulse. The injection laser can be focused to a very small spot size to decrease the transverse ionization region and due to its shorter Rayleigh length it will have a shorter $L_{inj}$. This leads to much reduced $\Delta\xi_\perp$ and $\Delta\xi_\gamma$. In a relativistic beam driver case, the phase velocity of the nonlinear wake is equal to the velocity of the driver bunch, which is typically closer to the speed of light than the group velocity of the laser. Therefore the term $\Delta\xi_\gamma$ is much reduced. The electron is longitudinally frozen in the wake after it is boosted to relativistic energy (it does not dephase). This longitudinal position can therefore be defined as $\xi_f$ and this is insensitive when it was ionized.

\begin{figure}[bp]
\includegraphics[width=0.1666\textwidth]{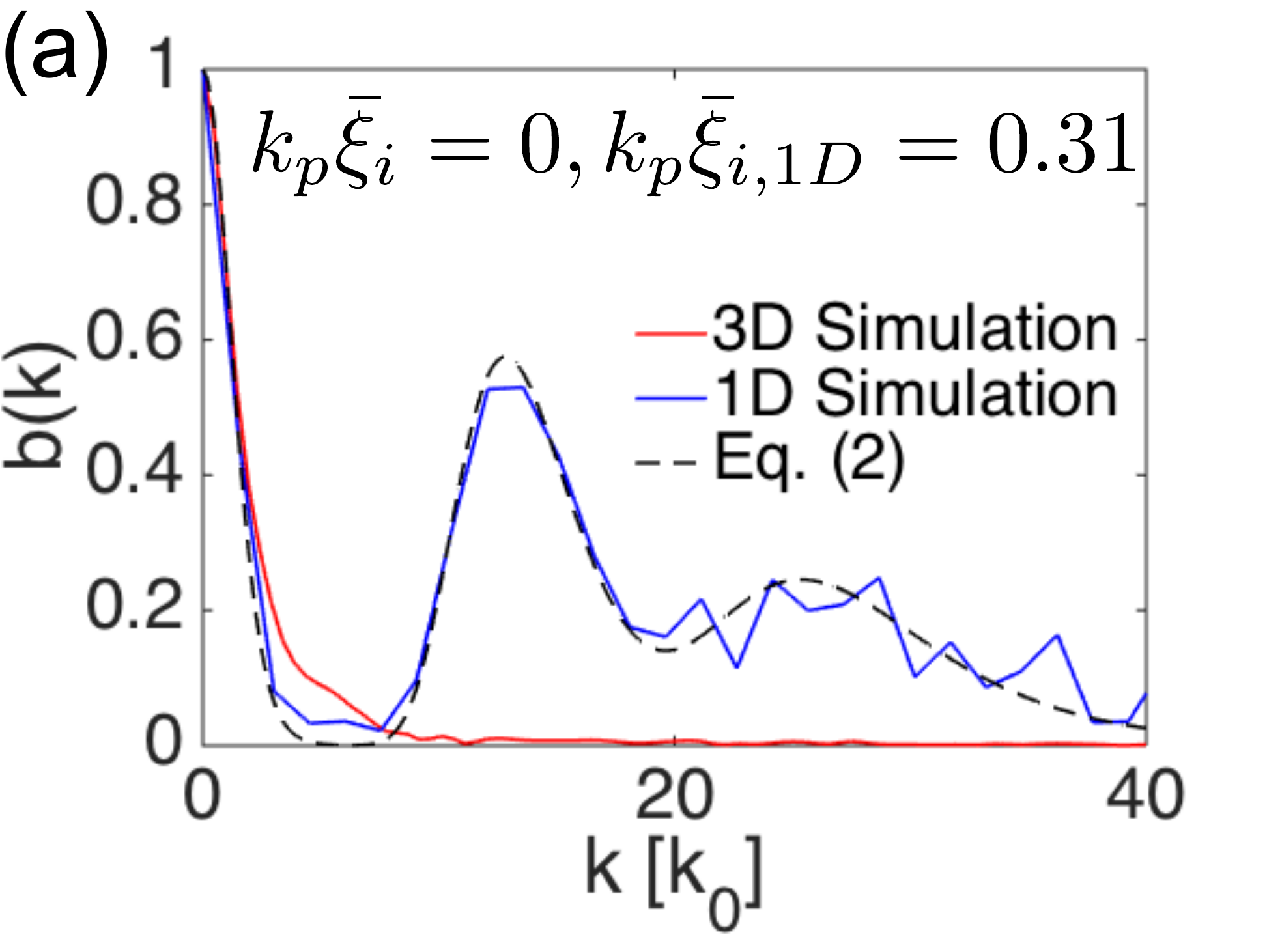}\includegraphics[width=0.1666\textwidth]{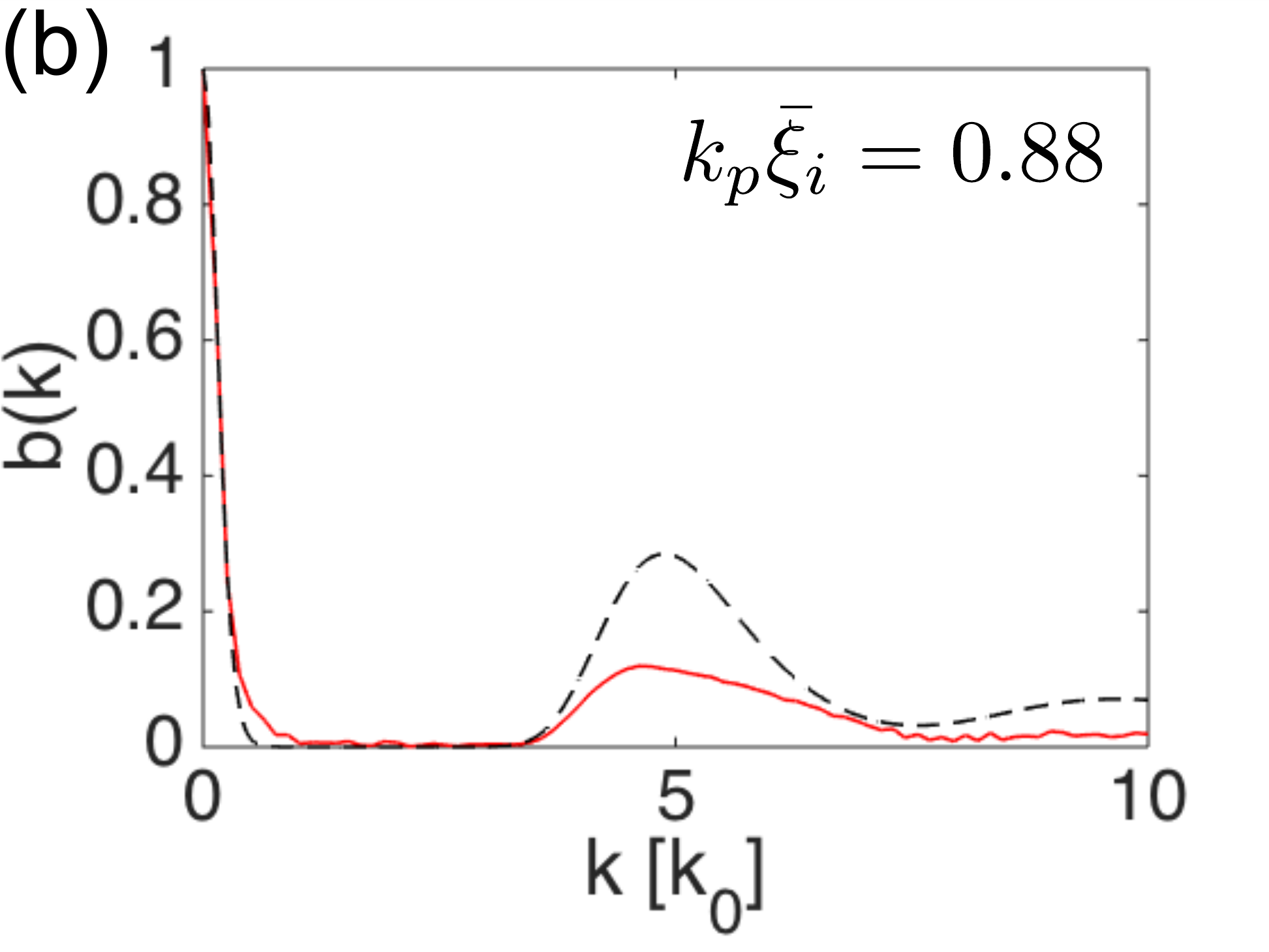}\includegraphics[width=0.1666\textwidth]{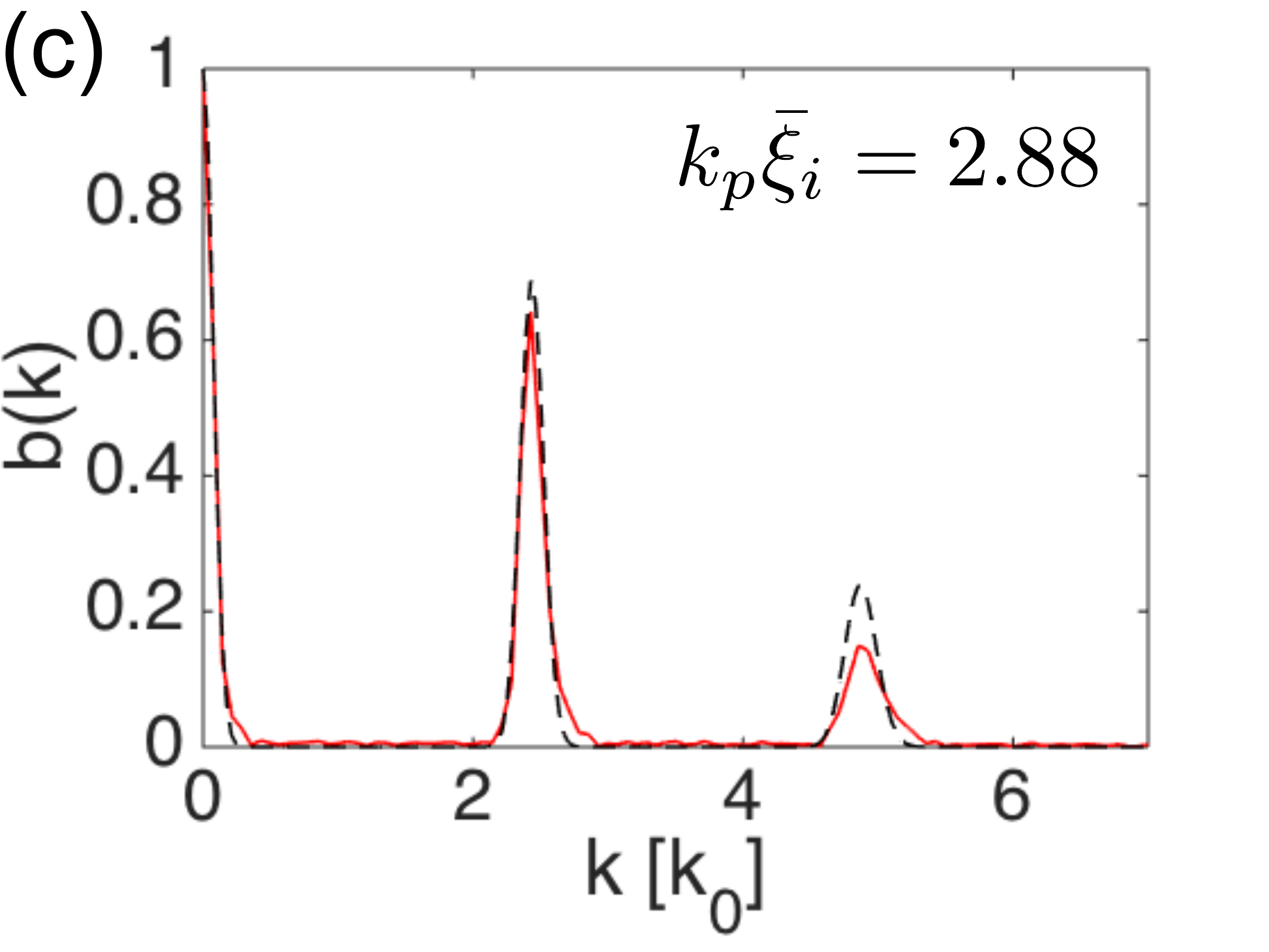}
\includegraphics[width=0.1666\textwidth]{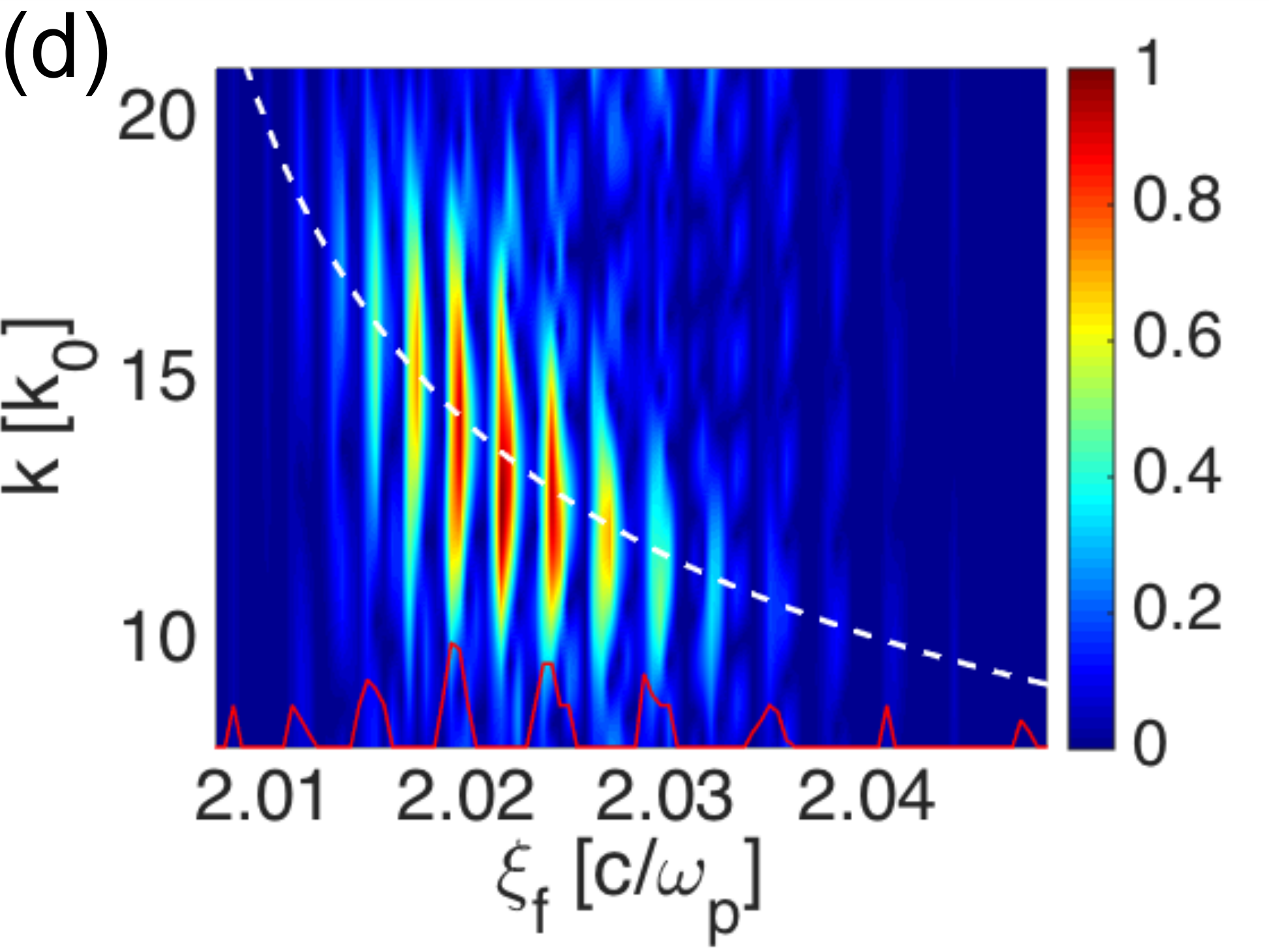}\includegraphics[width=0.1666\textwidth]{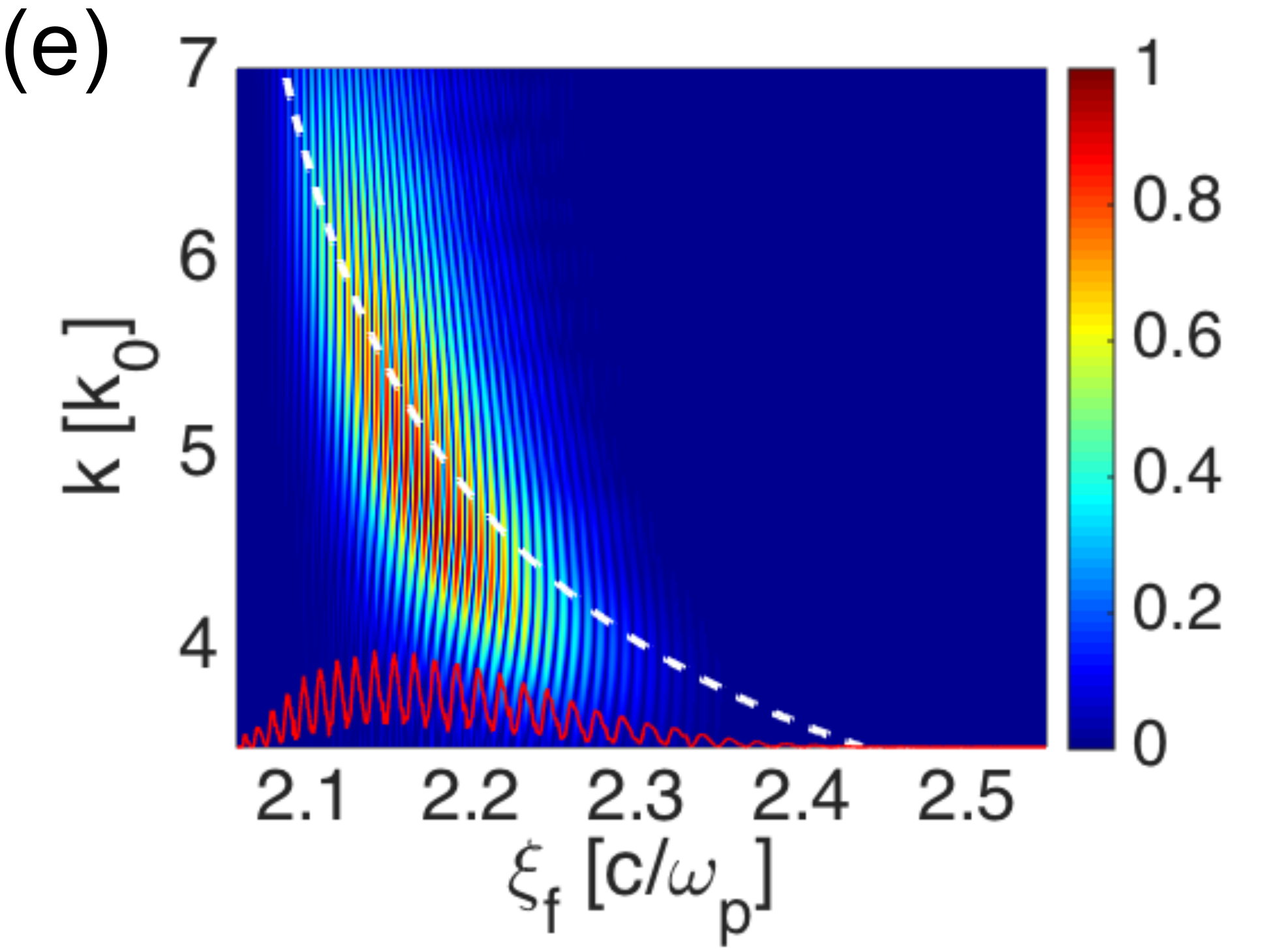}\includegraphics[width=0.1666\textwidth]{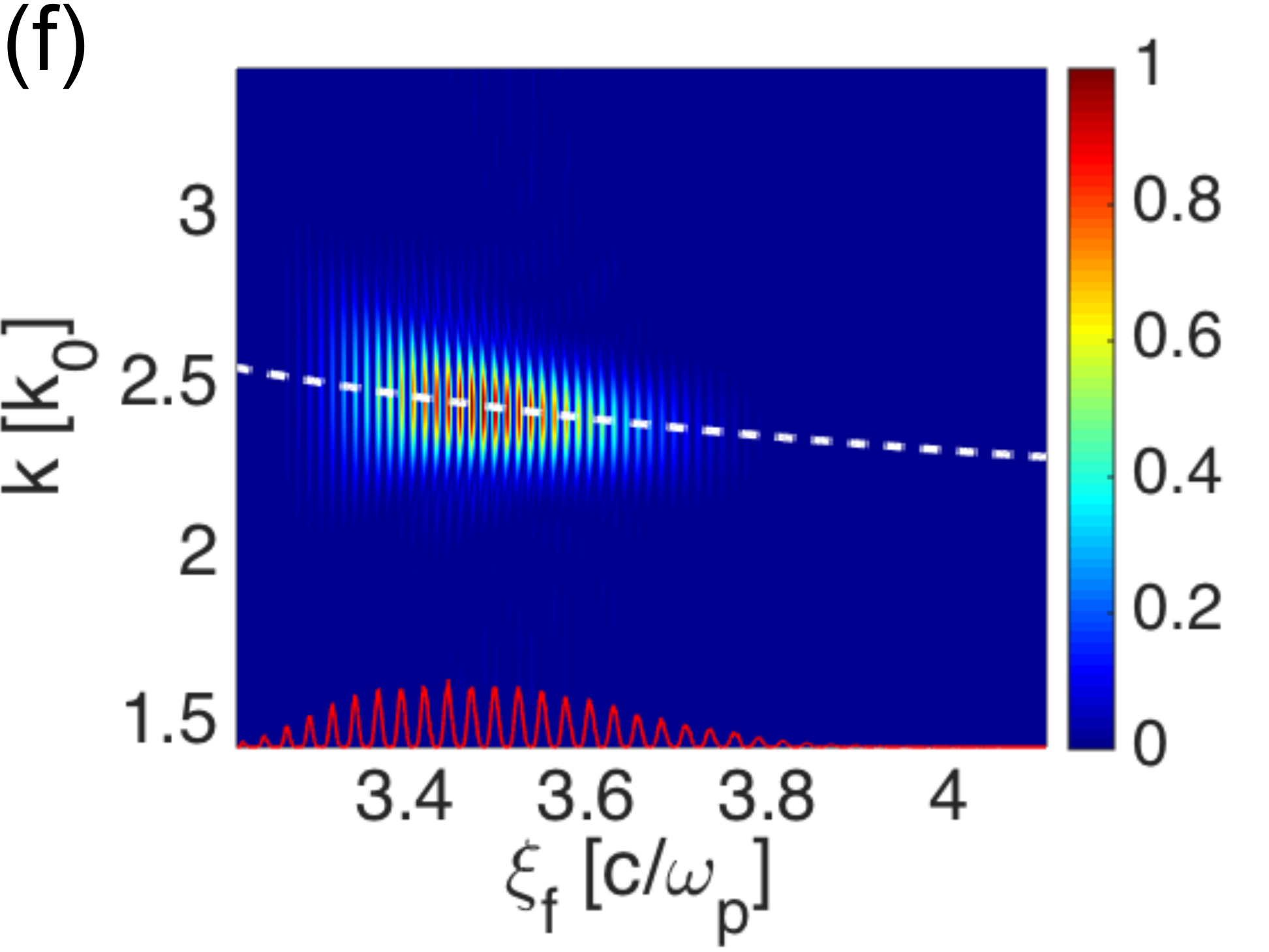}
\caption{\label{fig: } Bunching of electrons when charges is injected by using a relatively low intensity laser pulse in prescribed and non evolving wakefield. (a), (b) and (c) The red lines are the bunching factor while scanning the initial $\bar{\xi}_i$ and the dashed lines are the analytical results from Eq. 2. (d), (e) and (f) The corresponding Wigner transform of the current profile showing the spatial modulation. The white dashed lines are $k/k_0=2\sqrt{4+\xi_i^2}/\xi_i$ and the red lines are the current profile of the injected charge. Note that (d) is 1D simulation and others are 3D simulations.}
\end{figure}

Under the assumption of no phase slippage in the wake, the effect of the nonlinear mapping between $\xi_i$ and $\xi_f$ and the finite $r_i$ on the bunching factor can be quantified as follows. The distribution of the final longitudinal parameters $\xi_f$ is $g(\xi_f)d\xi_f = d\xi_i\int dr_i f(r_i, \xi_i)$ can be obtained from the distribution of the initial parameters is $f(\xi_i,r_i)$, where $r_f$ is neglected which is also reasonable when the energy of the electron is high. The bunching factor is $b(k)=\left| \int d \xi_f \mathrm{exp}(ik\xi_f) g(\xi_f)\right| = \left| \int d \xi_i dr_i \mathrm{exp}(ik\xi_f) f(\xi_i,r_i)\right| $. Assuming $\delta\xi_i \equiv \xi_i-\bar{\xi}_i \ll \bar{\xi}_i$ and $r_i \ll \bar{\xi}_i$ where $\bar{\xi}_i$ is the mean value of $\xi_i$, then after expanding $\xi_f$ to the order of $O(r_i^2)$ and $O(\delta\xi_i^2)$, $\xi_f$ can be expressed as $\xi_f\approx\sqrt{4+\bar{\xi}_i^2}\left[ 1+ \frac{\delta\xi_i}{h_m^2\bar{\xi}_i} + \frac{\delta\xi_i^2}{2h_m^2\bar{\xi}_i^2}\left(1-\frac{1}{h_m^2}\right) + \frac{r_i^2}{2h_m^2\bar{\xi}_i^2}\right] $, where $h_m={\sqrt{4+\bar{\xi}_i^2}}/{\bar{\xi}_i}$ is the wavenumber upshift factor obtained from the nonlinear mapping process. We assume the initial distribution is $f(\xi_i,r_i) = \frac{r_i}{ \sigma_r^2}\mathrm{exp}\left( -\frac{r_i^2}{2\sigma_r^2}\right) \frac{1}{\sqrt{2\pi}\sigma_e}\mathrm{exp}\left(-\frac{\delta\xi_i^2}{2\sigma_e^2}\right)  \sum\limits_{n=-\infty}^{+\infty}F_{n}\mathrm{exp}(-i2nk_0 \delta\xi_i) $, where $F_n = \int d\xi_i f_b(\xi_i)\mathrm{exp}(i2nk_0\delta\xi_i)$ and $f_b(\xi_i)$ is the initial $\xi_i$ distribution in a single slice. Substitute the expression of $f(\xi_i, r_i)$ into the bunching factor, then it is straightforward to obtain 
\begin{align}
b(k)=  \sum_{n=-\infty}^{+\infty} \left| F_n\right| R(\hat{\sigma}_r, \hat{\sigma}_e) \mathrm{e}^{ -\frac{(k - 2n h_m k_0)^2 \sigma_e^2}{2 h_m^2 (1+\hat{\sigma}_e^4)}} \label{eq: b}
\end{align}
where $R(\hat{\sigma}_r, \hat{\sigma}_e)=\left[(1+\hat{\sigma}_r^4)(1+\hat{\sigma}_e^4)\right]^{-1/4}$ is the 3D reduction factor, $\hat{\sigma}_{e}=\sigma_{e}\sqrt{(1-1/h_m^2)k/(h_m \bar{\xi}_i)}$ and $\hat{\sigma}_{r}=\sigma_{r}\sqrt{k/(h_m \bar{\xi}_i)}$. The ratio of the strongest modulation wavenumber in the current profile over the wavenumber of the injection laser (the modulation factor) is
\begin{align}
h=2h_m={2\sqrt{4+\bar{\xi}_i^2}}/{\left| \bar{\xi}_i \right|} \label{eq: harmonic number}
\end{align}
where the factor $2$ is from the ionization process and the factor $h_m$ is from the nonlinear mapping process. Eq. (3)  shows that the wavelength of the modulation is shortest for $\bar{\xi}_i$ near zero (near the maximum of the wake potential). However, $\hat{\sigma}_r$ becomes very large for $\bar{\xi}_i$ near zero, therefore, from Eq. (2), $R$ will be small in this limit. For this reason the wave number of the modulation is limited and the modulation is only seen when ionization occurs off the maximum of the potential.

\begin{figure}[bp]
\includegraphics[width=0.5\textwidth]{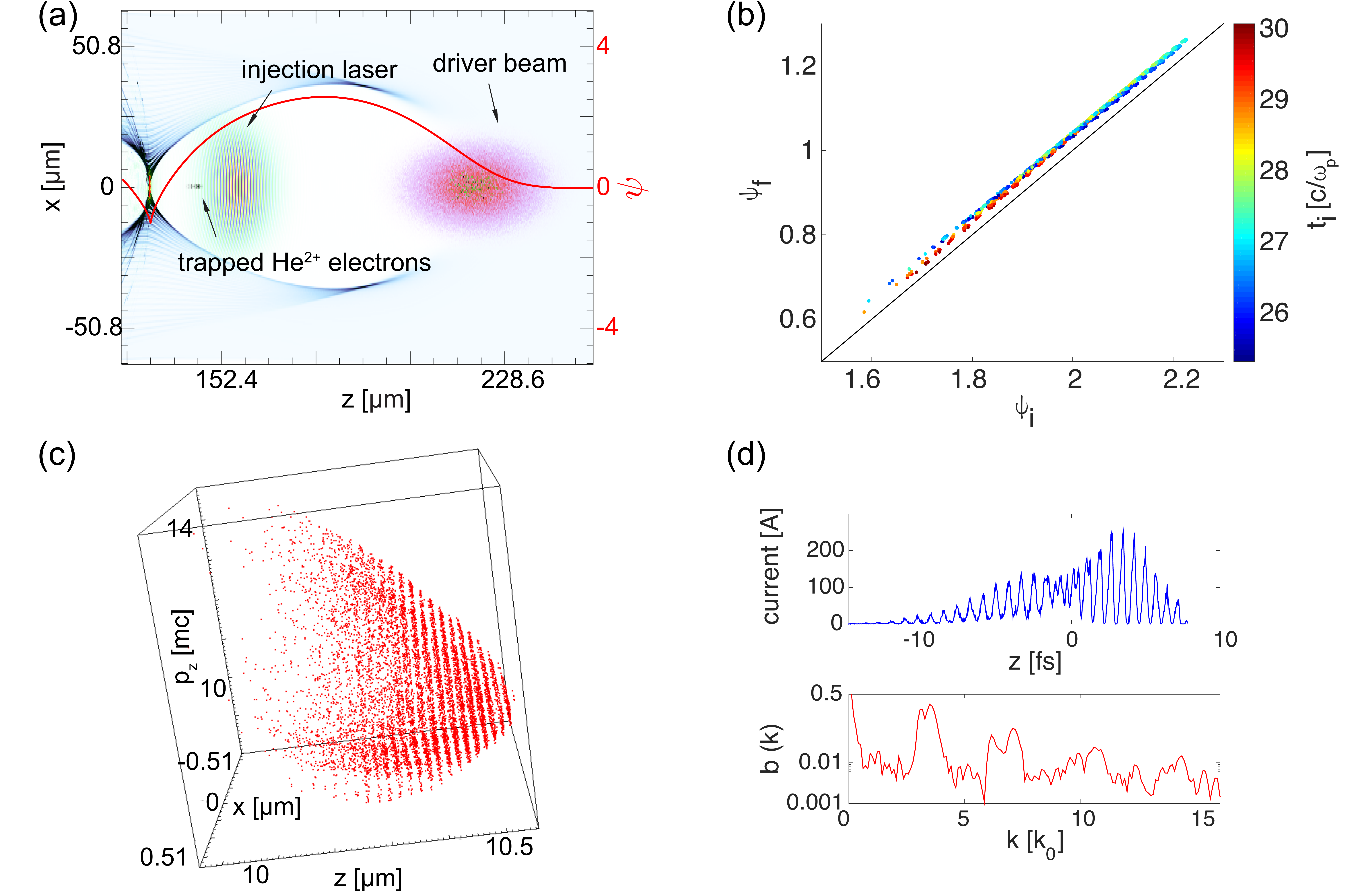}
\caption{\label{fig: } An electron driver-beam followed by an injection laser propagate to the right in a mixture of pre-ionized plasma and He$^{1+}$ ions ($n_p=1.74\times 10^{17}~\centi\meter^{-3}, n_{\mathrm{He}^{1+}} = 0.1 n_p$). Driver-beam: $E_b = 1~\giga\electronvolt, \sigma_r=8.9~\micro\meter, \sigma_z = 10.6~\micro\meter, I_b=19~\kilo\ampere$; The injection laser is as same as in the case of Fig. 2. (a). Snapshot of the charge density distribution of the background electrons, the 2nd electron of helium, and the laser electric field. The red line is the pseudo potential at the center. (b) The dependence of the final $\psi_f$ on the initial $\psi_i$ for ionization injected electrons. The color represents the ionization time. The black line represents $\psi_f-\psi_i=-1$. (c) The $\left(z,x,p_z\right)$ phase space distribution of the trapped electrons at $z =0.1~\milli\meter$. (d) The modulated ($h\approx 3$) current profile and the bunching factor of the trapped electrons at $z =0.1~\milli\meter$. }
\end{figure}

These conclusions are verified numerically. We use OSIRIS with non-evolving forces from the nonlinear wakefields, i.e., $F_z=-\xi/2, F_r=r/2 + (1-v_z)r/2$. An 800 nm laser with $a_0=0.12, w_0=2~\micro\meter$ propagates through a plasma with $n_p=1.74\times 10 ^{17}~\centi\meter^{-3}$ and a $10^{-5}n_p$ $\mathrm{He}^{1+}$ plasma (to minimize space charge effects) provides the ionized electrons. The longitudinal delay between the laser pulse and the plane with $\xi=0$ is scanned to generate electrons with different $\bar{\xi}_i$ and the resulting bunching factors are shown in Figs. 2(a)-(c). When the laser is strongly focused, $\gamma-p_z-\psi$ is not strictly conserved and the variation leads to a reduction of the bunching factor which more serious when $h$ is larger [see Fig. 2(b)]. Due to the nonlinear mapping process, the modulation factor depends on $\xi_f$, which can be seen from the Wigner transform of the current profile as shown in Figs. 2(d)-(e). The modulation factor can be very high theoretically when $\bar{\xi_i}$ is very close to zero, but in this case $R$ is very small so $b$ is rather small. However in a 1D simulation, $h$ as high as $15$ was observed as shown in Fig. 2 (d).

\begin{figure}[bp]
\includegraphics[width=0.25\textwidth]{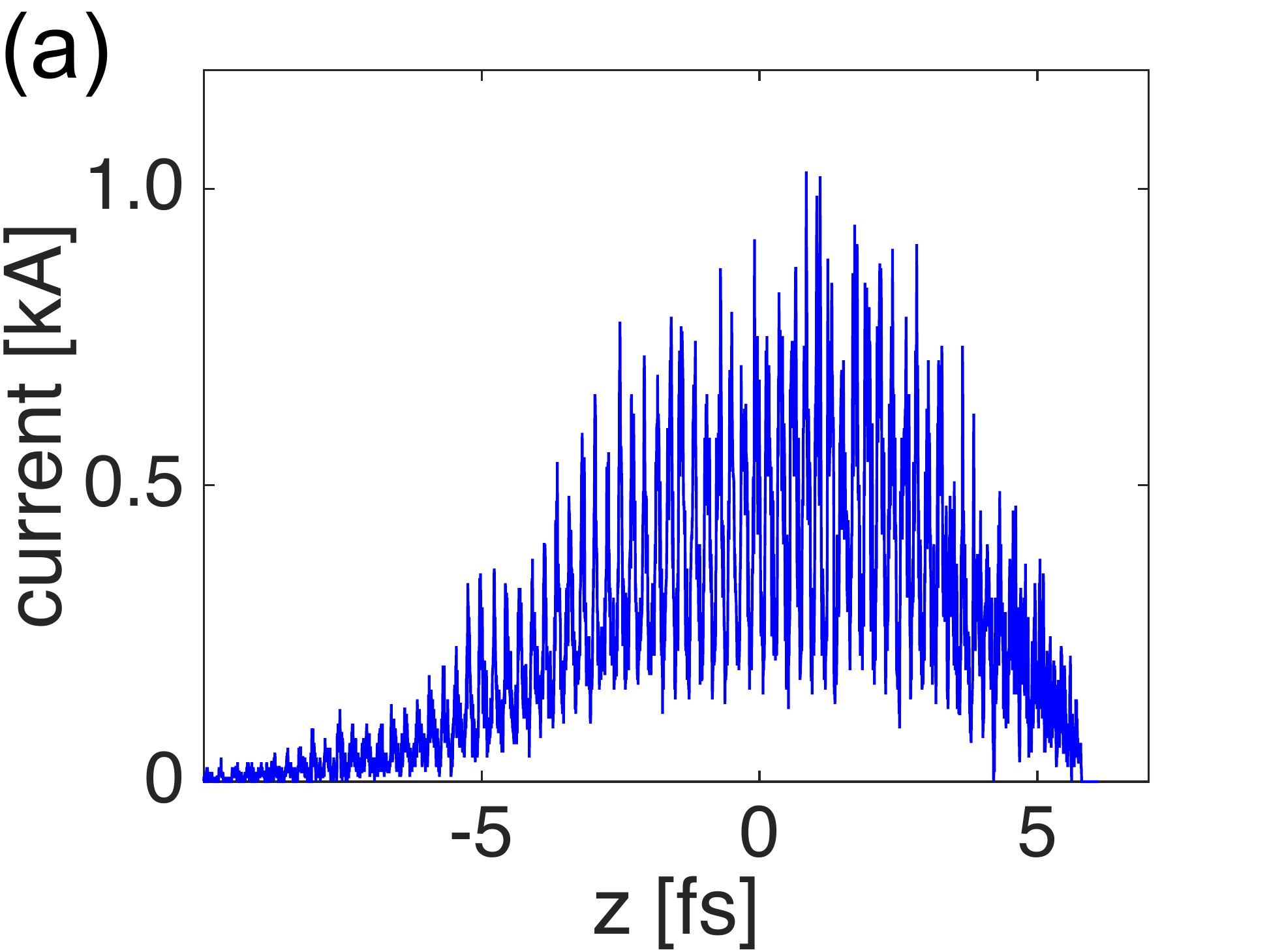}\includegraphics[width=0.25\textwidth]{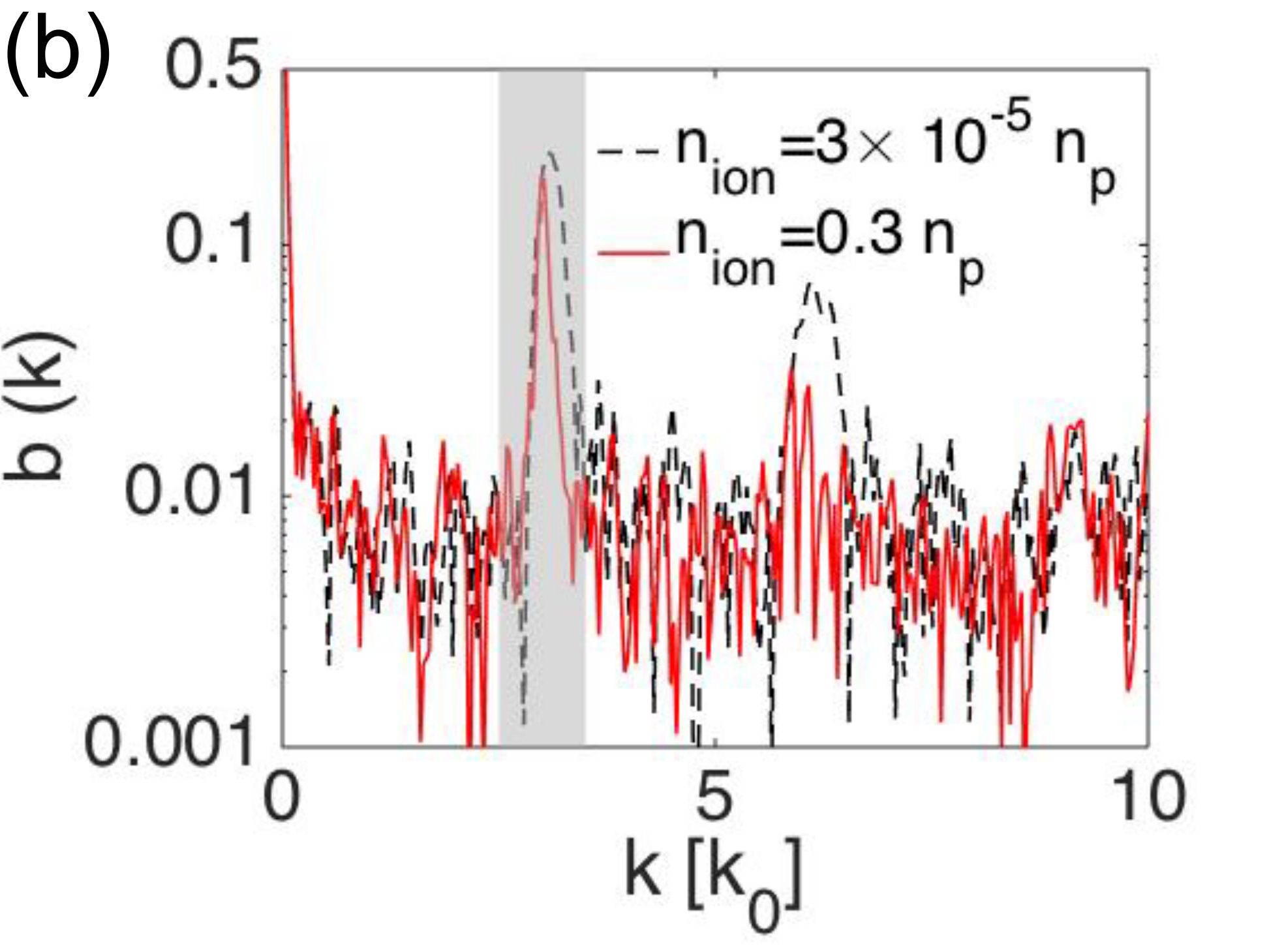}
\caption{\label{fig: } The current profile (a) and bunching factor (b) of the trapped electrons by using 200 $\nano\meter$ injection laser at $z=0.1~\milli\meter$. The red lines are the results when $n_{\mathrm{He}^{1+}} = 0.3~ n_p$ and the black dashed line is the result when $n_{\mathrm{He}^{1+}} = 3\times10^{-5}~ n_p$.}
\end{figure}

We next present results from a fully self-consistent 3D OSIRIS simulation. We use a relativistic electron beam to produce the wake for the ionization of the inner shell electrons. A mixture of preionized plasma and He$^{1+}$ ions is used. The electric field of the electron beam is low enough to not doubly ionize the helium. The simulation window has a dimension of $127\times 127 \times 127 ~\micro\meter$ with $1000\times 1000\times 5000$ cells in the $x, y$ and $z$ directions, respectively. This corresponds to cell sizes of $k_0^{-1}$ in the $x$ and $y$ directions and 0.2 $k_0^{-1}$ in the $z$ direction. There are 4 particles per cell to represent the $\mathrm{He}^{1+}$ ions. An 800 $\nano\meter$ injection laser with the same amplitude, spot size and pulse length used above (Fig. 2) is focused into the wake as shown in Fig. 3(a). The laser is focused at $z=0~\milli\meter$ while the plasma starts from $z=-0.254~\milli\meter$. By tracking particles, we confirm that the trapping condition \cite{PhysRevLett.104.025003} $\psi_f\approx \psi_i - 1$ as shown in Fig. 3(b). In Fig. 3(c) we present the phase space distribution of the trapped charge after $z=0.1~\milli\meter$ for a case where the relative longitudinal position between the beam driver and the injection laser is chosen to achieve $\bar{\xi}_i\approx1.87$. For this case, based on Eq. 3 the predicted modulator factor is $h\approx 2.9$. The current profile of the electron beam and the bunching factor $b(k)$ at this distance are shown in Fig. 3(d). 

By replacing the 800 $\nano\meter$ injection laser by its 4th harmonic - 200 $\nano\meter$ injection laser, an electron bunch with a strong UV frequency modulation is generated. We simulate this using OSIRIS with the external wakefield model described above for the same plasma density. The $\mathrm{He}^{1+}$ density was either $3\times 10^{-5}n_p$ or $0.3n_p$. A 200 $\nano\meter$ injection laser with $a_0=0.023, w_0=2~\micro\meter$ are used to release the 2nd electron of helium. The current profile and the bunching factor at $z=0.1~\milli\meter$ are shown in Figs. 4(a) and (b) respectively where it is seen that individual electrons are micro bunched spatially on a nano-scale (attosecond in the time domain). Such a micro-bunched structure will give rise to intense coherent transition \cite{leemans2003observation} at the harmonics of the bunching frequency. The coherent transition radiation energy generated from a sharp plasma-vacuum boundary generated by the modulation within  the shaded wave number (frequency) shown in Fig. 4(b) is about 0.1 nJ \cite{landau1984electrodynamics} (we assume the beam has a mean energy $\bar{\gamma}=1000$). The radiated spectrum will contain the fundamental and the second harmonic of the nano-structured beam at 65.6 $\nano\meter$ and 32.8 $\nano\meter$ respectively. Detection of this coherent radiation at wavelengths shorter than the ionizing laser wavelength is a good diagnostic of this self-bunching in the wake. Space charge interaction between the injected electrons will blur the modulation at $nh k_0$ and thus reduce the modulation and the bunching factor at $nhk_0$ which can be seen from the comparison between the dashed line ($n_{\mathrm{He}^{1+}}=3\times10^{-5}n_p$) and the solid line ($n_{\mathrm{He}^{1+}}=0.3 n_p$) in Fig. 4(b). 

Due to the small spot size and low intensity of the injection laser, the emittance and energy spread of the trapped beam are both very small, e.g, for the 200 $\nano\meter$ injection laser case discussed above $\epsilon_{nx}=10.9~\nano\meter, \epsilon_{ny}=10.6~\nano\meter$, and $\sigma_\gamma = 3.2$. If such an electron beam can be accelerated further, extracted from the plasma and coupled into a short, resonant undulator without degrading its emittance \cite{xu2014coherent} it will produce intense coherent radiation. For example, consider an electron beam with $\bar{\gamma} = 1068.9$ and $\sigma_\gamma = 3.2$ propagating into a planar undulator with wavelength $\lambda_u=3~\centi\meter$ and normalized undulator parameter $K =2 $. The undulator is resonant at the modulation wavelength of the electron beam, $\lambda_r= 65.6~\nano\meter$. The output radiation power saturates at $P_{sat} = 400~\mega\watt$ in 3 $\meter$ undulator when by simulating this process with 3D  GENSIS 1.3 code \cite{reiche1999genesis}. 

In conclusion, we have shown that the discrete injection of the electrons due to the laser ionization injection process is mapped to the final phase space of the accelerated beam in a plasma accelerator operating in the blowout regime. Theoretical analysis and 3D PIC simulations are presented. This intrinsic phase space discretization phenomenon leads to nano-scale micro bunching of the accelerated beam that can be diagnosed through coherent transition radiation upon the beam's exit from the plasma and may find use in generating high-power ultraviolet radiation upon passage through a resonant undulator. 

Work supported by NSFC grants 11175102, 11005063, thousand young talents program, DOE grants DE-SC0010064, DE-SC0008491, DE-SC0008316, and NSF grants ACI-1339893, PHY-1415386, PHY-0960344. Simulations are performed on the UCLA Hoffman 2 and Dawson 2 Clusters, and the resources of the National Energy Research Scientific Computing Center.

\bibliography{refs_xinlu}

\end{document}